\documentclass{aa}  

\usepackage{graphicx}
\usepackage{txfonts}
\usepackage{booktabs,hyperref}
\usepackage{microtype}
\usepackage[modulo,switch]{lineno}
\usepackage{natbib}

\newcommand{\uHz}{\hbox{$\mu$Hz}}
\newcommand{\st}[1]{_{\rm #1}}
\newcommand{\numax}{\nu\st{max}}
\newcommand{\Dnu}{\Delta\nu}
\newcommand{\Teff}{T\st{eff}}
\newcommand{\Msun}{\hbox{M$_\sun$}}
\newcommand{\Rsun}{\hbox{R$_\sun$}}
\newcommand{\Myr}{\hbox{Myr}}
\newcommand{\Gyr}{\hbox{Gyr}}
\newcommand{\eq}[1]{\begin{equation}#1\end{equation}}
\newcommand{\tabspread}{1ex}

\begin{document} 
\linenumbers

\title{Grid-based seismic modelling at high and low signal-to-noise ratios}
\subtitle{HD~181420 and HD~175272}
\author{S.~Hekker\inst{1} \and W.~H.~Ball\inst{2}}
\institute{Max-Planck-Institut f\"ur Sonnensystemforschung, Justus-von-Liebig-Weg 3, 37077 G\"ottingen, Germany\\
\email{hekker@mps.mpg.de}\and
Institut f\"ur Astrophysik, Georg-August-Universit\"at G\"ottingen, Friedrich-Hund-Platz 1, 37077 G\"ottingen, Germany}

\date{Received ; accepted }

\abstract
{Recently, the CoRoT target HD~175272 (F5V), which shows a weak signal of solar-like oscillations, 
was modelled by a differential asteroseismic analysis (Ozel et al. 2013) relative to a seismically similar star, HD~181420  (F2V),
for which there is a clear signature of solar-like oscillations. The results provided by Ozel et al. (2013) indicate the possibility of HD~175272 having subsolar mass, while being of the order of 1000~K hotter than the Sun. This seems unphysical -- standard stellar evolution theory generally does not predict solar-metallicity stars of subsolar mass to be hotter than about 6000K -- and calls for a reanalysis of this star.}
{We aim to compare the performance of differential asteroseismic analysis with that of grid-based modelling. }
{We use two sets of stellar model grids and two grid-fitting methods to model HD~175272 and HD~181420  using their effective temperatures, metallicities, large frequency separations, and frequencies of maximum oscillation power as observational constraints.}
{We find that we are able to model both stars with parameters that are both mutually compatible
and comparable with other modelling efforts. Hence, with modest spectroscopic and asteroseismic inputs, we obtain reasonable estimates of stellar parameters. In the case of HD~175272, the uncertainties of the stellar parameters from our grid-based modelling are smaller, and hence more physical, than those reported in the differential analysis.
For both stars, the models have significantly lower values of $\numax$ than the 
reported observed values. 
Furthermore, when using the asymptotic large frequency separation as opposed to the scaling relation to compute $\Dnu$, we find that our modelling results are significantly more self-consistent when $\numax$ is ignored. }
{Grid-based modelling is a useful tool even in cases of weak solar-like oscillations. It provides more precise and more realistic results than obtained with differential seismology. The difference in the observed and modelled values of $\numax$ indicates that the four observational constraints are not fully consistent with 
the stellar models used here, with $\numax$ most likely to be the inconsistent constraint for these two stars.
}


\maketitle
%

\section{Introduction}
The CoRoT \citep{baglin2006,auvergne2009} and \textit{Kepler} \citep{borucki2008,borucki2010} missions have provided a wealth of high-quality, nearly-uninterrupted photometric time series, which are useful for investigations of stellar oscillations. From stellar oscillations in main-sequence stars, subgiants, and red giants it is possible to accurately derive stellar parameters such as mass, radius, mean density, and surface gravity, as well as more detailed knowledge about the internal stellar structure. For example, it has been possible to measure: 
whether \emph{helium is burning} in the cores of red giants \citep{beck2011,bedding2011,mosser2011mm}; 
how much (differential) rotation takes place \citep{beck2012,deheuvels2012,mosser2012rot};
the inclination angle and obliquity between the \emph{spin axes of a star and its companions} \citep{chaplin2013obl,huber2013};
the location of the second helium ionization zone \citep{miglio2010,mazumdar2012,mazumdar2014}.

Solar-like oscillations -- oscillations stochastically excited in the turbulent outer convective zones of low mass main-sequence stars, subgiants, and red giants -- can be analysed to various levels of detail. Accurate measures of individual frequencies are important to study the internal structures of stars. However, it is not always possible to measure individual oscillation frequencies owing, for example, to a low signal-to-noise ratio of the oscillations. In such cases one can use so-called global oscillation parameters: the frequency separation between modes of the same degree and consecutive (acoustic) radial order $\Delta\nu$, which is roughly proportional to the square root of the mean density of the star; and the frequency of maximum oscillation power $\numax$, which scales with the surface gravity $g$ and square root of the effective temperature $\Teff$. These two parameters can be used to estimate the stellar mass and radius, when combined with effective temperature using well-known scaling relations \citep{kjeldsen1995}.

The scaling relations are useful but should always be used with their limitations in mind. 
Firstly, they can be derived by assuming homology in stellar structures. 
This is clearly an approximation because structural features like the depths of convection zones and the presence (or absence) of convective cores change with mass, metallicity, and age. 
Comparisons with independently-derived radii and surface gravities have shown that the approximation is reasonably accurate for a wide range of stars, although with an increase in uncertainty for stars with different internal stellar structures compared to Sun-like stars \citep[e.g.][]{white2011,huber2012,silva2012,miglio2012}.  Secondly, the $\Delta\nu$ scaling relation is derived in the asymptotic regime where the radial order $n$ is much larger than the degree $l$, i.e. $n\gg l$. This is usually true for oscillations of stars on the main sequence, but not necessarily for more evolved stars. Corrections to account for the non-asymptotic regime are currently being discussed \citep{mosser2013as,hekker2013as}.

To constrain stellar age and composition, one can compare the observed global oscillation parameters with those of a large number of stellar models. This is often referred to as \emph{grid-based} modelling \citep[e.g.][]{basu2010,gai2011}. It has found wide application mostly in the analysis of large sets of data \citep[e.g.][]{chaplin2014,hekker2013}, but also of stars with weak signatures of oscillation \citep[e.g.][]{barclay2013}.

Recently, \citet{ozel2013} proposed to derive stellar parameters for a star with a weak oscillation signal by computing linear differences with respect to the stellar parameters in a second star with a strong oscillation signal and similar observable parameters (e.g. $\Dnu$, $\numax$, $T_{\rm eff}$, [Fe/H]).  
\citet{ozel2013} referred to this method as \emph{differential seismology} and applied it to the seismically similar stars HD~175272  (F5V) and HD~181420  (F2V) \citep{bruntt2009,barban2009,huber2012} using CESAM models.  HD~175272 was observed by CoRoT for 27 days and shows oscillations at a low signal-to-noise ratio.  Though the individual oscillation frequencies cannot be identified for HD~175272, global seismic parameters can be determined and were found to be similar to the parameters of the well-studied star HD~181420. This star was therefore used as a reference to derive the parameters of HD~175272. 

In this work we present grid-based modelling for both HD~175272 and HD~181420, using two different sets of stellar models and two different grid-modelling methods. The observational constraints for HD~181420 and HD~175272 were taken from \citet{ozel2013} and are listed in Table \ref{input}.  We compare our results with the results from differential asteroseismology.

\begin{table}
\begin{minipage}{\linewidth}
\caption{Observed parameters of HD~181420 (F2V) and HD~175272  (F5V) taken from \citet{ozel2013}.}
\label{input}
\centering
\begin{tabular}{lcc}
\toprule
& HD~181420 & HD~175262 \\
\midrule
$\Dnu/\uHz$     & 75.20 $\pm$ 0.04 &  74.9 $\pm$  0.4 \\
$\numax/\uHz$   &  1610 $\pm$ 10   &  1600 $\pm$   30 \\
$\Teff/{\rm K}$ &  6580 $\pm$ 100  &  6675 $\pm$  120 \\
$\rm[Fe/H]$     & $-$0.05 $\pm$ 0.06 & +0.08 $\pm$ 0.11 \\
\bottomrule
\end{tabular}
\end{minipage}
\end{table}

\begin{table*}
\caption{Grid-based modelling results.}
\label{bigtab}
\begin{tabular}{l@{\hskip1cm}cccc@{\hskip1cm}cccc}
\toprule
           & \multicolumn{4}{c}{HD~181420} & \multicolumn{4}{c}{HD~175272} \\
Model grid & BASTI & BASTI & MESA\tablefootmark{a} & MESA\tablefootmark{b}
           & BASTI & BASTI & MESA\tablefootmark{a} & MESA\tablefootmark{b} \\
Method     & LD & PySEEK & PySEEK & PySEEK 
           & LD & PySEEK & PySEEK & PySEEK \\
\midrule
$M/M_\sun$ & 
$1.40\pm0.04$ & $1.40^{+0.03}_{-0.03}$ & $1.41^{+0.04}_{-0.02}$ & $1.49^{+0.05}_{-0.02}$ &   
$1.48\pm0.08$ & $1.48^{+0.06}_{-0.07}$ & $1.50^{+0.07}_{-0.07}$ & $1.59^{+0.05}_{-0.07}$ \\ [\tabspread]  
$t/\text{Gyr}$ & 
$1.81^{+0.07}_{-0.07}$ & $1.8^{+0.5}_{-0.1}$& $2.0^{+0.2}_{-0.4}$& $1.3^{+0.2}_{-0.2}$ &
$1.4^{+1.0}_{-0.6}$ & $1.3^{+0.5}_{-0.3}$& $1.4^{+0.4}_{-0.4}$& $1.0^{+0.4}_{-0.3}$\\ [\tabspread]
$R/R_\sun$& 
$1.65\pm0.06$ & $1.65^{+0.01}_{-0.02}$& $1.66^{+0.02}_{-0.01}$ & $1.712^{+0.002}_{-0.002}$ &
$1.68\pm0.06$ & $1.69^{+0.02}_{-0.03}$& $1.69^{+0.03}_{-0.03}$ & $1.74^{+0.01}_{-0.03}$\\ [\tabspread]
$\Delta\nu/\uHz$& 
$75.24\pm0.01$& $75.26^{+0.05}_{-0.16}$& $75.20^{+0.19}_{-0.03}$& $75.16^{+0.04}_{-0.01}$ &
$75.1\pm0.5$ & $75.1^{+0.4}_{-0.4}$& $75.1^{+0.4}_{-0.4}$& $75.1^{+0.4}_{-0.4}$\\ [\tabspread]
$\numax/\uHz$& 
$1483.5\pm 0.4$ & $1484^{+2}_{-3}$  & $1490^{+4}_{-1}$ & $1449^{+30}_{-11}$ &
$1496\pm25$ & $1500^{+17}_{-20}$& $1508^{+17}_{-24}$& $1484^{+27}_{-24}$\\ [\tabspread]
$\Teff/{\rm K}$ &
$6574^{+14}_{-14}$& $6583^{+7}_{-165}$ & $6552^{+124}_{-72}$ & $6780^{+175}_{-82}$ &
$6662^{+147}_{-144}$ & $6648^{+153}_{-87}$ & $6661^{+125}_{-116}$ & $6810^{+116}_{-119}$ \\ [\tabspread]
$\rm[Fe/H]$ &
$-0.25 \pm 0.01$  & $0.06^{+0.09}_{-0.09}$ & $0.06^{+0.09}_{-0.09}$ & $0.06^{+0.09}_{-0.09}$ &
$0.07 \pm 0.03$ & $0.22^{+0.11}_{-0.18}$ & $0.20^{+0.11}_{-0.18}$ & $0.25^{+0.09}_{-0.15}$ \\
\bottomrule
\end{tabular}
\tablefoot{\tablefoottext{a}{$\Delta\nu$ from scaling relations.}
\tablefoottext{b}{$\Delta\nu$ from asymptotic large frequency spacing ($\Dnu\st{as}$, see Eq. \ref{nu_as})}.}
\end{table*}

\begin{table*}
\caption{Parameters derived for HD~181420 in the literature.}
\label{litHD18}
\centering
\begin{tabular}{lcccccc}
\toprule
\tiny{reference} & \tiny{\citet{bruntt2009}} & \tiny{\citet{mathur2010}}\tablefootmark{a} & \tiny{\citet{mathur2010}}\tablefootmark{b} & \tiny{\citet{huber2012}} & \tiny{\citet{ozel2013}}\tablefootmark{c} & \tiny{\citet{ozel2013}}\tablefootmark{d}\\
\midrule
$M/\Msun$ & 1.31 $\pm$ 0.06 & 1.6 $\pm$ 0.3 & & 1.6 $\pm$ 0.2 & 1.3 $\pm$ 0.2 & 1.3 $\pm$ 0.2\\
$t/\Gyr$ & 2.7 $\pm$ 0.4 & & & & 2.1 $\pm$ 0.2 & 2.3 $\pm$ 0.3 \\
$R/R_\sun$ & 1.60 $\pm$ 0.03 & 1.7 $\pm$ 0.2 & 1.61 $\pm$ 0.03 & 1.73 $\pm$ 0.08 & 1.6 $\pm$ 0.1 & 1.6 $\pm$ 0.1\\
\bottomrule
\end{tabular}
\tablefoot{\tablefoottext{a}{scaling relations}\tablefoottext{b}{Radius Extractor \citep{creevey2013}}\tablefoottext{c}{GN93}\tablefoottext{d}{AGS05}}
\end{table*}

\section{Grid-based modelling}

\subsection{Stellar model grids}
Our grid-based modelling uses two stellar model grids.  The first is the canonical BASTI grid\footnote{\url{http://albione.oa-teramo.inaf.it/}} \citep{basti2004}, which spans masses from 0.5~$\Msun$ to 3.5~$\Msun$ in steps of 0.05~$\Msun$ and metallicities of $Z=0.0001$, $0.0003$, $0.0006$, $0.001$, $0.002$, $0.004$, $0.008$, $0.01$, $0.0198$, $0.03$ or $0.04$.  (The corresponding helium abundances are $Y=0.245$, $0.245$, $0.246$, $0.246$, $0.248$, $0.251$, $0.256$, $0.259$, $0.2734$, $0.288$, and $0.303$.) The BASTI grid includes models from the zero-age main sequence all the way to the asymptotic giant branch phase.  The models were computed using an updated version of the code described by \citet{cassisi1997} and \citet{salaris1998}.  Opacities were taken from the OPAL tables \citep{iglesias1996} and \citet{ferguson2005} for $\log T>4$ and $\log T\leq4$, respectively.  The solar abundance pattern and solar metallicity were taken from \citet{grevesse1993} and no diffusion was included.  Nuclear reaction rates are drawn from the NACRE database \citep{angulo1999}, with the particular rate for ${}^{12}\mathrm{C}(\alpha,\gamma){}^{16}\mathrm{O}$ from the calculation by \citet{kunz2002}.  The models used a contemporary version of the FreeEOS equation of state\footnote{\url{http://freeeos.sourceforge.net/}} \citep{irwin2012} and the atmosphere model by \citet{krishna-swamy1966}.  Convection is described by the formulation of mixing-length theory \citep{boehm-vitense1958} given by \citet{cox1968}, with the mixing-length parameter fixed at a solar-calibrated value.  Finally, mass loss is included according to{the empirical relation by \citet{reimers1975} with the efficiency parameter $\eta$ taking the widely-used value of 0.4.

The second grid was constructed using MESA\footnote{\url{http://mesa.sourceforge.net/}} \citep[][revision 5527]{paxton2011,paxton2013} and covered the same selection of initial metallicities and helium abundances as the BASTI grid but with masses{from 0.8~$\Msun$ to 1.8~$\Msun$ in steps of 0.01~$\Msun$.  Models were evolved from the pre-main-sequence up to a maximum age of 15 Gyr or a maximum radius of 25~$\Rsun$, well up the red-giant branch, with a maximum timestep of 50~$\Myr$.  Opacities are drawn from the OPAL tables \citep{iglesias1996} and \citet{ferguson2005} at high and low temperatures, respectively, and are smoothly blended over the entire region where the tables overlap ($3.75\leq\log T\leq4.50$).  The models use a standard Eddington grey atmosphere and convective processes are described by mixing-length theory according to \citet{henyey1965} with mixing-length parameter $\alpha=1.9$.  All models use a scaled solar composition according to the values given by \citet{grevesse1998} without any chemical diffusion.  Nuclear reaction rates are taken from \citet{caughlan1988} or the NACRE collaboration \citep{angulo1999}, with preference given to the latter when available.  Unlike the BASTI grid, no mass loss was included.  All other options took their default values described in the MESA instrument papers \citep{paxton2011,paxton2013}.

\subsection{Asteroseismic parameters}
Given the mass $M$, radius $R$, and effective temperature $\Teff$ of a stellar model, $\numax$ and $\Dnu$ can be estimated using the commonly-used scaling relations \citep{kjeldsen1995}:
\eq{\frac{\numax}{\nu_{{\rm max},\sun}}=\frac{M}{\Msun}\left(\frac{R}{\Rsun}\right)^{-2}\left(\frac{\Teff}{T_{{\rm eff},\sun}}\right)^{-\frac{1}{2}},}
and
\eq{\frac{\Dnu}{\Dnu_\sun}=\left(\frac{M}{\Msun}\right)^\frac{1}{2}\left(\frac{R}{\Rsun}\right)^{-\frac{3}{2}},}
where the subscript $\sun$ denotes solar values, for which we took
$\nu_{{\rm max},\sun}=3090\pm 30\,\uHz$, $\Dnu_\sun=135.1\pm0.1\,\uHz$, and $T_{{\rm eff},\sun}=5778\,{\rm K}$ \citep{huber2011}. The uncertainties in the solar values are taken into account in the grid-based modelling by adding them quadratically to the uncertainties in the observations \citep{chaplin2014}. We note here that we did not use the corrections to the observed and reference $\Delta\nu$ values proposed by \citet{mosser2013as} as the analysis of a large sample of solar-like stars has shown that this does not have significant impact on the results \citep{chaplin2014}.

For the MESA grid, we also computed the asymptotic large frequency separation $\Dnu\st{as}$ given by
\eq{\Dnu\st{as}=\left(2\int_0^{R}\frac{dr}{c(r)}\right)^{-1},\label{nu_as}}
where $c$ is the local speed of sound and $r$ the radial co-ordinate inside the star.  In other words,
the asymptotic large frequency separation is the inverse of the time taken for a sound wave to travel from one 
side of the star to the other.  
The MESA grid includes the model values of $\Dnu\st{as}$, so we performed two separate fits: one with $\Dnu$ from the scaling relations and one with $\Dnu$ equal to its asymptotic value.  These are marked by superscript \emph{a} and \emph{b} in Table \ref{bigtab}.
When using the asymptotic separation, we did not include the uncertainty in $\Dnu_\sun$ in the observational constraints.

\subsection{Model fitting methods}
We used two methods to fit stellar models to observations. The first method is an independent implementation of the likelihood method described by \citet{basu2010}. In short, the likelihood of each model is computed given the values of some chosen set of observed parameters. To obtain a reliable uncertainty for the derived parameters a Monte Carlo analysis is performed, in which the observed values are perturbed within their uncertainties and a new likelihood is determined. The final answer is derived from the centre and width of a Gaussian fit through the total likelihood distribution of 1000 perturbations. We refer to this method as LD (Likelihood Distribution).

The second method is an independent implementation of the SEEK method \citep{quirion2010}, which we refer to as PySEEK.  PySEEK calculates the Gaussian probability of each stellar model given the observations and constructs a probability-weighted histogram for the desired parameter.  The cumulative sum of the histogram is computed and a linear interpolation is used to find where the cumulative sum takes values $0.1585$, $0.5$, and $0.8415$, which correspond to the one-sigma limits of a Gaussian distribution.  These values are then taken as the $\pm1\sigma$ intervals that are reported.  This is the same method that was used for the \textsc{GOE} results reported in \citet{chaplin2014}.

\begin{figure}
\begin{minipage}{\linewidth}
\centering
\includegraphics[width=\linewidth]{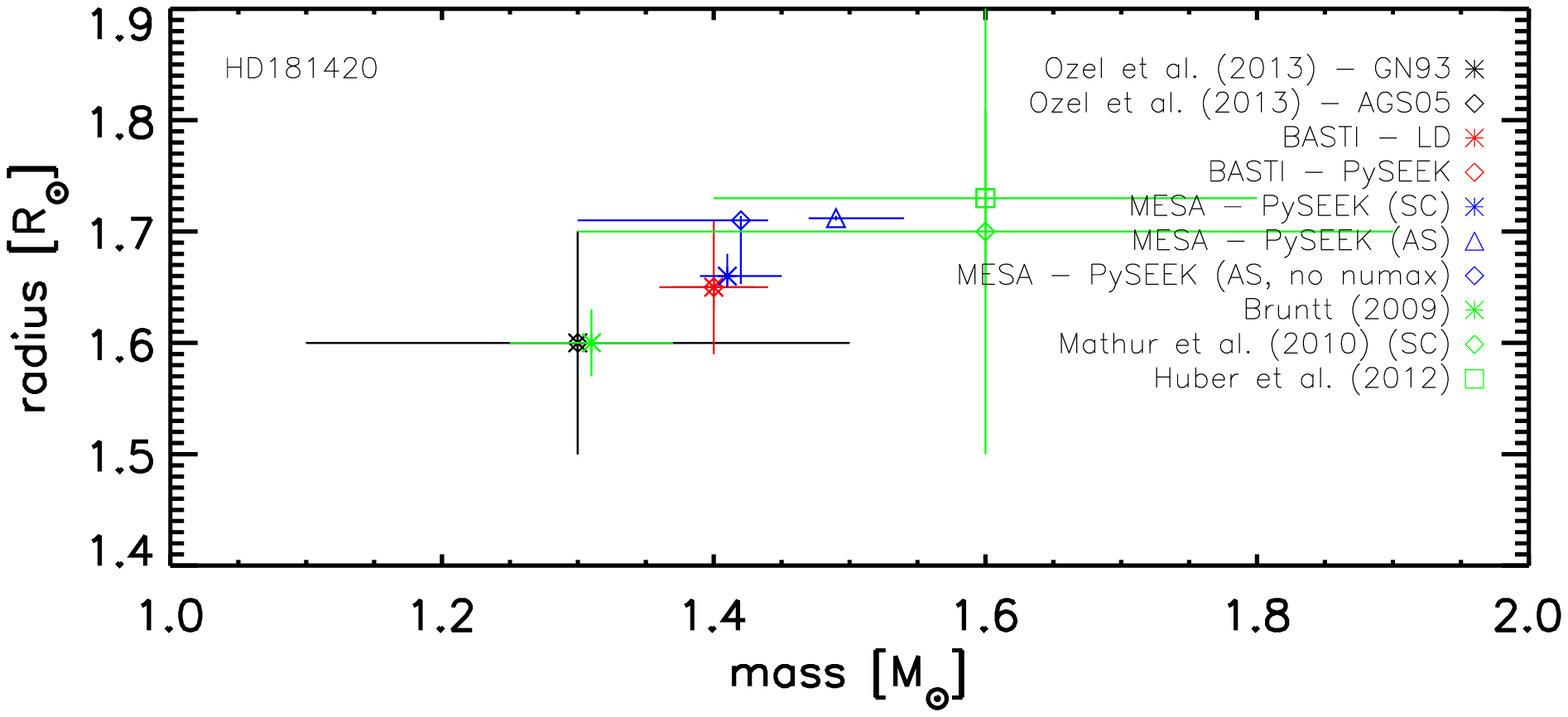}
\end{minipage}
\hfill
\begin{minipage}{\linewidth}
\centering
\includegraphics[width=\linewidth]{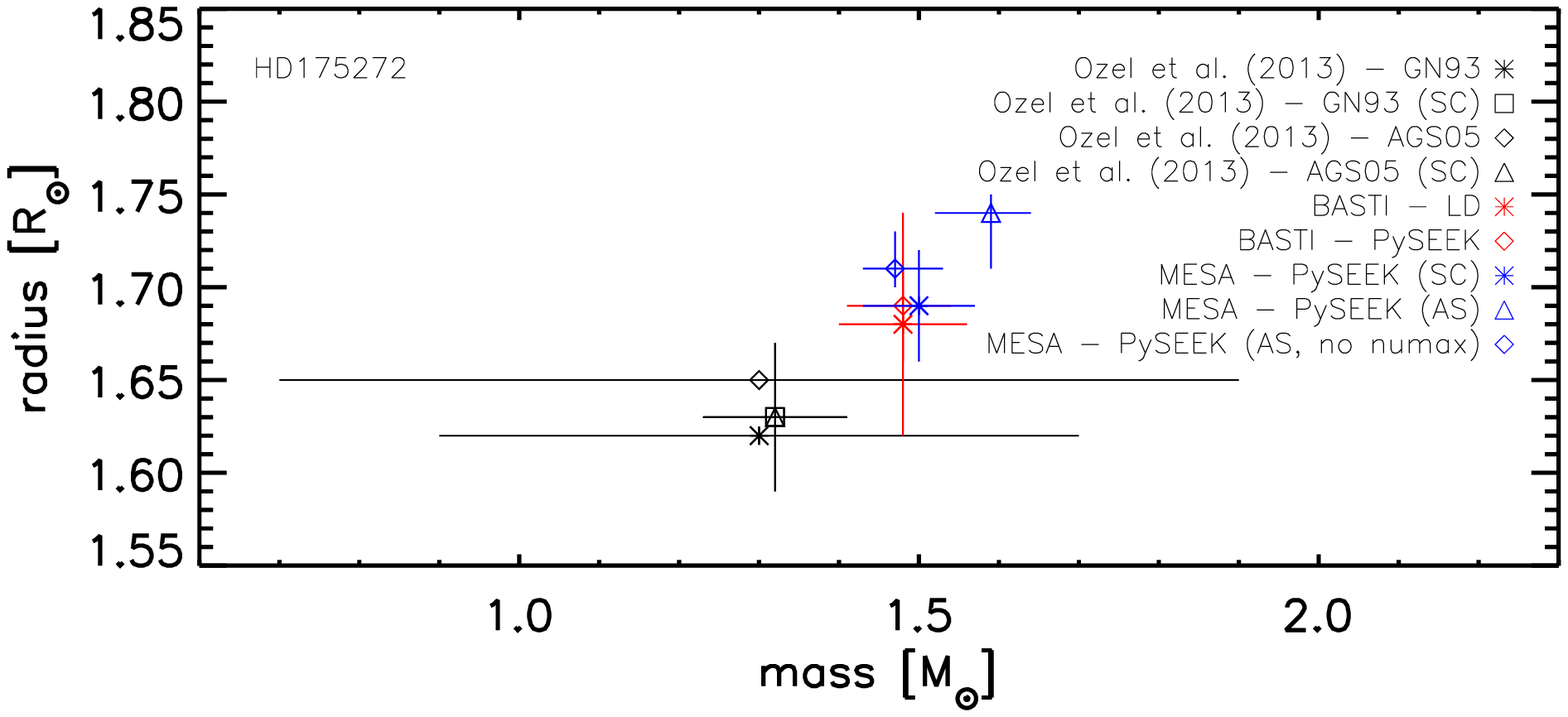}
\end{minipage}
\caption{Radii and masses determined for HD~181420 (top) and HD~175272 (bottom) using the differential seismology using CESAM models (black, Tables~\ref{litHD18} and \ref{litHD17}), grid-based modelling using BASTI models (red, Table~\ref{bigtab}), grid-based modelling using MESA models (blue, Tables~\ref{bigtab} and \ref{MESA2}), and literature values (green, Table~\ref{litHD18}). See legend for more information.}
\label{MRplots}
\end{figure}

\section{Results}

We present the results of our grid-based modelling in Table~\ref{bigtab}. Fig.~\ref{MRplots} shows fitted values of the masses and radii, along with results from other works.  The four sets of results are mutually compatible. When both using the BASTI models,  the two different grid search methods agree very well, as should be expected. 
The results derived using scaling relations and the MESA grid also agree with both sets of BASTI-based results.  When fitting $\Delta\nu$ with the asymptotic large separation (Eq.~\ref{nu_as}), the results for the MESA grid differ and consistently show larger masses, radii, and effective temperatures.
The differences in the masses and radii are consistent with the differences noted by \citet{white2011} in the sense that the models that are fitted with the asymptotic separation have smaller mean densities.  Though the parameters for HD~175272 still agree reasonably with the results derived with scaling relations, the parameters for HD~181420 are at best marginally consistent, with the radius being tightly constrained by the small uncertainty in the large frequency separation.
It remains unclear why the discrepancy is so significant, although we found that the discrepancy is eliminated when only $\Teff$, [Fe/H] and $\Dnu$ are used as observational constraints (see Section \ref{discussion} and Fig.~\ref{MRplots}).

Although the model results are mutually consistent, 
the values of $\numax$ returned from the grid-based modelling are all systematically lower than the reported observed value. We have checked the literature for additional values for $\numax$. For HD~181420 the following results are mentioned: $1574\pm13$~$\mu$Hz \citep{huber2012} and $1536.26\pm0.42$~$\mu$Hz \citep{mathur2010}. These values are intermediate between the observational value quoted by \citet{ozel2013} and the results from our grid-based modelling.

\begin{table}
\begin{minipage}{\linewidth}
\caption{Parameters for HD~175272 derived by \citet{ozel2013}.}
\label{litHD17}
\centering
\begin{tabular}{lcccc}
\toprule
mixture & \multicolumn{2}{c}{GN93} & \multicolumn{2}{c}{AGS05} \\
$\Dnu$ & scaling & adiabatic & scaling & adiabatic \\
       & rel. & freq. & rel. & freq. \\
\midrule
$M/\Msun$ & $1.32\pm0.09$ & $1.3\pm0.4$ & $1.32\pm0.09$ & $1.3\pm0.6$ \\
$t/\Gyr$ & $1.6\pm0.3$ & $1.6\pm0.6$ & $1.8\pm0.2$ & $1.7\pm0.1$ \\
$R/R_\sun$ & $1.63\pm0.04$ & $1.62$ & $1.63\pm0.04$ & $1.65$ \\
\bottomrule
\end{tabular}
\end{minipage}
\end{table}

\section{Discussion}
\label{discussion}
HD~181420 has been extensively studied with several techniques. \citet{bruntt2009} performed a detailed spectroscopic analysis of the star followed by an in-depth seismic analysis by \citet{barban2009}. \citet{huber2012} obtained an independent radius measurement using interferometry. This star was further studied by \citet{mathur2010} and \citet{ozel2013}. The parameters reported in these works are listed in Table~\ref{litHD18}. The masses and radii are roughly consistent with our results. The ages, where given, are slightly (and not always significantly) larger. We note that these results all correspond to lower masses. To achieve the same radius (which is strongly constrained by asteroseismology), lower-mass stars must be older, so we expect a correlation between the ages and masses. Additionally, the results obtained by \citet{ozel2013} are derived using $\Delta\nu$, $T_{\rm eff}$, and the luminosity $L$, while in the current analysis luminosity was not constrained, which could also influence the determined mass and age.

Being a lower signal-to-noise target, HD~175272 has been less extensively studied.  \citet{ozel2013} modelled HD~175272 using a differential analysis and their results are listed in Table~\ref{litHD17}.  These are based on two solar mixtures (\citealt{grevesse1993}, GN93; \citealt{asplund2005}, AGS05) and two calculations of the large separation: either from the scaling relations or from a direct calculation of the adiabatic oscillation frequencies.  Again, our results are broadly consistent. However, the uncertainties quoted by \citet{ozel2013} predict that HD~175272 could have a subsolar mass, whereas its \emph{effective} temperature is about 1000~K hotter than the Sun's. Based purely on the effective temperature of 6675 $\pm$ 120 K, and given that metallicity of HD 175272 is roughly solar, we would expect subsolar masses to be excluded at around the 5-sigma level. The uncertainties in our grid-based modelling are significantly smaller, and do not predict such unrealistic values. Thus, our results demonstrate that grid-based modelling can provide consistent and realistic stellar parameters even given only modest spectroscopic and asteroseismic constraints.

One may argue that the results presented here are only for the input physics implemented in the BASTI and MESA models, and thus the uncertainties quoted in Table~\ref{bigtab} do not include a contribution from uncertainties in the model physics. \citet{chaplin2014} performed a similar grid-based modelling analysis of a large sample of dwarfs and subgiant stars using 6 different grid-based pipelines and 11 stellar evolution codes. These authors conclude that, for stars for which spectroscopic data are available, as is the case for HD~175272 and HD~181420, the median uncertainties are approximately 5.4\% in mass, 2.2\% in radius and 57\% of their sample have ages with uncertainties less than 1 Gyr. These values are of the same order as the values quoted in Table~\ref{bigtab} and would not imply unphysically large uncertainties. We note here that only internal error bars are shown in Fig.~\ref{MRplots} as literature values only mention internal uncertainties.

We now return to the discrepancy between the observed and modelled values of $\numax$ \citep[see also][]{ozel2013}. For both stars all of our fits correspond to $\nu_\text{max}\lesssim1500\,\mu\text{Hz}$, which is significantly lower than the observed values of $\sim$$1600\,\uHz$. 
The modelled values for $\Dnu$, [Fe/H] and $T_{\rm eff}$ are consistent with the observed values, given the uncertainties and restrictions from the model grid. In particular, both grids are relatively coarse in metallicity
and the values returned are dominated by the closest possibilities compared to the observed values, with the uncertainties reflecting the width of the bins of either the likelihood distribution (LD) or the cumulative distribution function (PySEEK).

In addition, we found that by fitting the MESA models with the asymptotic large separations to the three constraints on $\Teff$, [Fe/H] and $\Dnu$ (and \emph{not} $\numax$), the discrepancies with the results from the scaling relations are completely eliminated.  The results of this fit are given in Table \ref{MESA2}. For both HD~181420 and HD~175272, the model parameters become much more consistent with the other results.  The values of $\numax$ become more discrepant, but all other fitted parameters move towards either the observed values or the values from the other models.
We conclude that $\numax$ is the discrepant observational constraint in HD~181420 and HD~175272.

\begin{table}
\begin{minipage}{\linewidth}
\caption{Results for grid-based modelling of HD~181420 with only $\Dnu$, $\Teff$, and [Fe/H] as observational constraints.}
\label{MESA2}
\centering
\begin{tabular}{l@{\hskip1cm}c@{\hskip1cm}c}
\toprule
           & HD~181420 & HD~175272 \\
Model grid & MESA\tablefootmark{b} & MESA\tablefootmark{b} \\
Method     & PySEEK & PySEEK \\
\midrule
$M/M_\sun$ & $1.42^{+0.02}_{-0.12}$ & $1.47^{+0.06}_{-0.04}$ \\ [\tabspread]
$t/\text{Gyr}$ & $2.0^{+0.6}_{-0.2}$ & $1.6^{+0.4}_{-0.3}$\\ [\tabspread]
$R/R_\sun$& $1.710^{+0.004}_{-0.057}$ & $1.71^{+0.02}_{-0.01}$\\ [\tabspread]
$\Delta\nu/\uHz$& $75.20^{+0.06}_{-0.01}$& $74.9^{+0.4}_{-0.4}$\\ [\tabspread]
$\numax/\uHz$& $1418^{+7}_{-48}$& $1431^{+28}_{-23}$\\ [\tabspread]
$\Teff/{\rm K}$ & $6582^{+72}_{-136}$ & $6666^{+125}_{-113}$ \\ [\tabspread]
$\rm[Fe/H]$ & $0.03^{+0.11}_{-0.13}$ & $0.08^{+0.01}_{-0.11}$ \\
\bottomrule
\end{tabular}
\tablefoot{
\tablefoottext{b}{$\Delta\nu$ from asymptotic large frequency spacing ($\Dnu\st{as}$, see Eq. \ref{nu_as}).}
}
\end{minipage}
\end{table}

\section{Conclusions}

We have successfully used grid-based modelling to derive consistent and realistic stellar parameters for both the high signal-to-noise target HD~181420 and the seismically similar star HD~175272.  Our analysis employed two model-fitting methods and two sets of stellar models. For HD~181420 we derive values in the ranges of 1.30--1.45~M$_{\odot}$, 1.59--1.714~R$_{\odot}$ and 1.7--2.6~Gyr in mass, radius, and age, respectively.
In the case of HD~175272, we derive smaller uncertainties than the differential analysis by \citet{ozel2013}, providing results in a physically expected regime, i.e., our results lie in the ranges 1.40--1.57 M$_{\odot}$, 1.62--1.73 R$_{\odot}$ and 1.27--1.64 Gyr for the mass, radius, and age, respectively. This demonstrates that grid-based modelling is useful even for low signal-to-noise targets.
We also note that the modelled values of $\Dnu$, $\Teff$ and [Fe/H] are consistent with the observed values given by \citet{ozel2013}, 
but the frequencies of maximum oscillation power $\numax$ are significantly smaller in both stars.  
In one modelling case, we found that omitting $\numax$ greatly improves the consistency of the models with the other observables.  We conclude that these observed values of $\numax$ are not consistent with the stellar models used here.

\begin{acknowledgements}
The authors would like to thank Jesper Schou, Hannah Schunker and Tim White for their comments on the manuscript.
WHB acknowledges research funding by Deutsche Forschungsgemeinschaft (DFG) under grant SFB 963/1 ``Astrophysical flow instabilities and turbulence'' (Project A18). The research leading to the presented results has received funding from the European Research Council under the European Community's Seventh Framework Programme (FP7/2007-2013) / ERC grant agreement no 338251 (StellarAges).
\end{acknowledgements}

\bibliographystyle{aa}
\bibliography{GBM}

\end{document}